\documentclass[11pt,oneside]{article}
\usepackage{amsfonts,amssymb,amsmath,verbatim,authblk,hyperref,graphicx,amsbsy,mathrsfs,caption,subcaption,cite,longtable}

\newcommand{\beq}{\begin{equation}\ }
\newcommand{\eeq}{\end{equation}\ }

\begin{document}

\author[1,2]{Demetris T. Christopoulos}
\affil[1]{\small{National and Kapodistrian University of Athens, Department of Economics}}
\affil[2]{\tt{dchristop@econ.uoa.gr}, \tt{dem.christop@gmail.com} }

\title{{\itshape Inflection point analysis for Greek seismicity 1900-2006}}

\maketitle

\begin{abstract}
The seismicity of Greek region is studied under the prism of ESE \& EDE methods for finding the inflection point of cumulative energy released. Main shocks are chosen from 106 years data. The result is that with both methods a critical time region exist at the end of 1982 to early 1983. After this time the seismicity tends to increase and gives remarkable events, like Athens Sep 1999 earthquake.\medskip
\end{abstract}


\section{Description of data used}
\begin{figure}
\begin{center}
\caption{Density plot of main events in Greece, 1900 - 2006}\label{fig:01}
\vspace{0.5in}  \includegraphics[width=11cm,height=11cm]{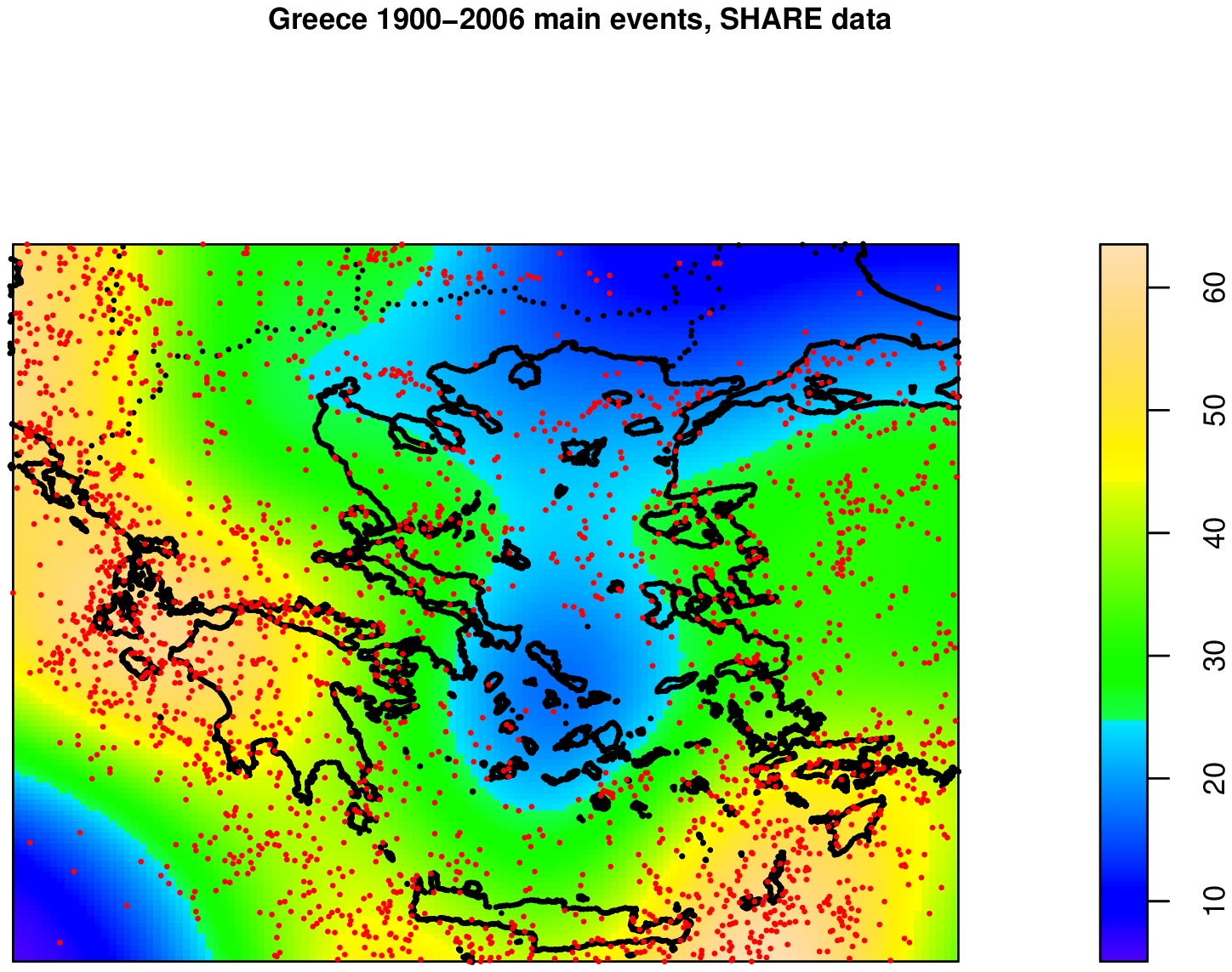}
\vspace{0.5in}
\end{center}
\end{figure}

\begin{figure}
\begin{center}
\caption{Counting of main events by region in Greece, 1900 - 2006}\label{fig:02}
\vspace{0.5in}  \includegraphics[width=13cm,height=13cm]{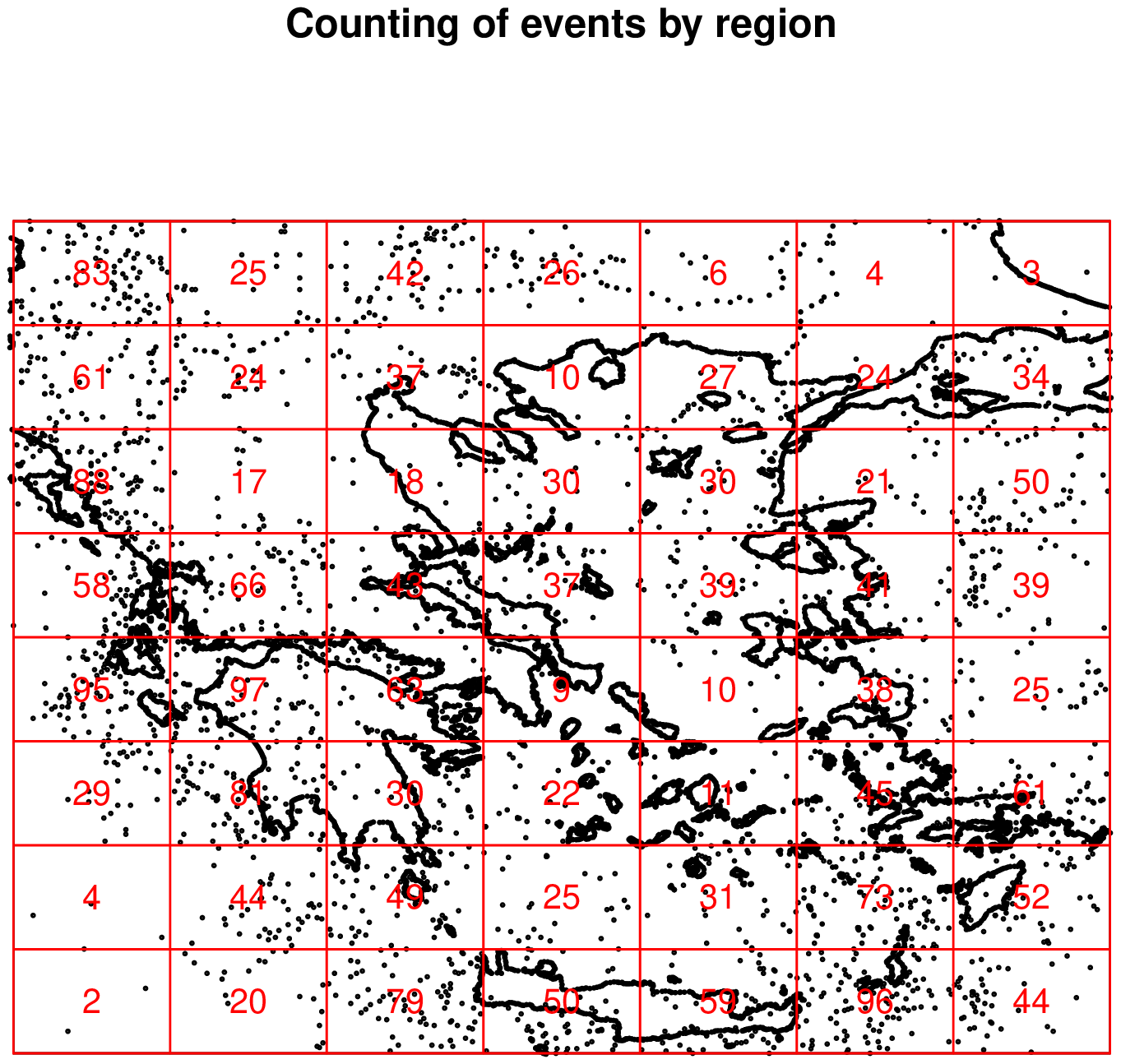}
\vspace{0.5in}
\end{center}
\end{figure}

We are using data from Seismic Hazard Harmonization in Europe project, see ~\cite{sha-13}, which has an index to show that an event  is a main one and not an aftershock one. We are interested to study only main events, so the above data is extremely useful. We have removed from data the entries without a valid magnitude value. We have also focused our data at the region of Longitude East $( 19.5000^{o},28.9984^{o})$
 and Latitude North $(34.8024^{o},41.9943^{o})$.\\
 The description of our data is presented at Table \ref{tab:01}. By using ~\cite{r-spatstat} we present the density plot of all main events at Figure \ref{fig:01}. The counting of events region by region is presented at Figure \ref{fig:02}. We clearly observe the main Greek seismicity arc where bigger events usually take place.

\begin{table}[!htbp] \centering 
   \caption{Descriptive statistics of Magnitude, main events, 1900-2006} 
    \label{tab:01} 
\begin{tabular}{@{\extracolsep{5pt}}lccccc} 
\\[-1.8ex]\hline 
\hline \\[-1.8ex] 
Statistic & \multicolumn{1}{c}{N} & \multicolumn{1}{c}{Mean} & \multicolumn{1}{c}{St. Dev.} & \multicolumn{1}{c}{Min} & \multicolumn{1}{c}{Max} \\ 
\hline \\[-1.8ex] 
Magnitude & 2,227 & 4.744 & 0.597 & 4.000 & 7.700 \\ 
\hline \\[-1.8ex] 
\normalsize 
\end{tabular} 
\end{table} 

\clearpage

\section{Inflection point analysis}
\begin{figure}
\begin{center}
\caption{Cumulative Magnitude in Greece, 1900 - 2006}\label{fig:03}
\vspace{0.5in}  \includegraphics[width=13cm,height=13cm]{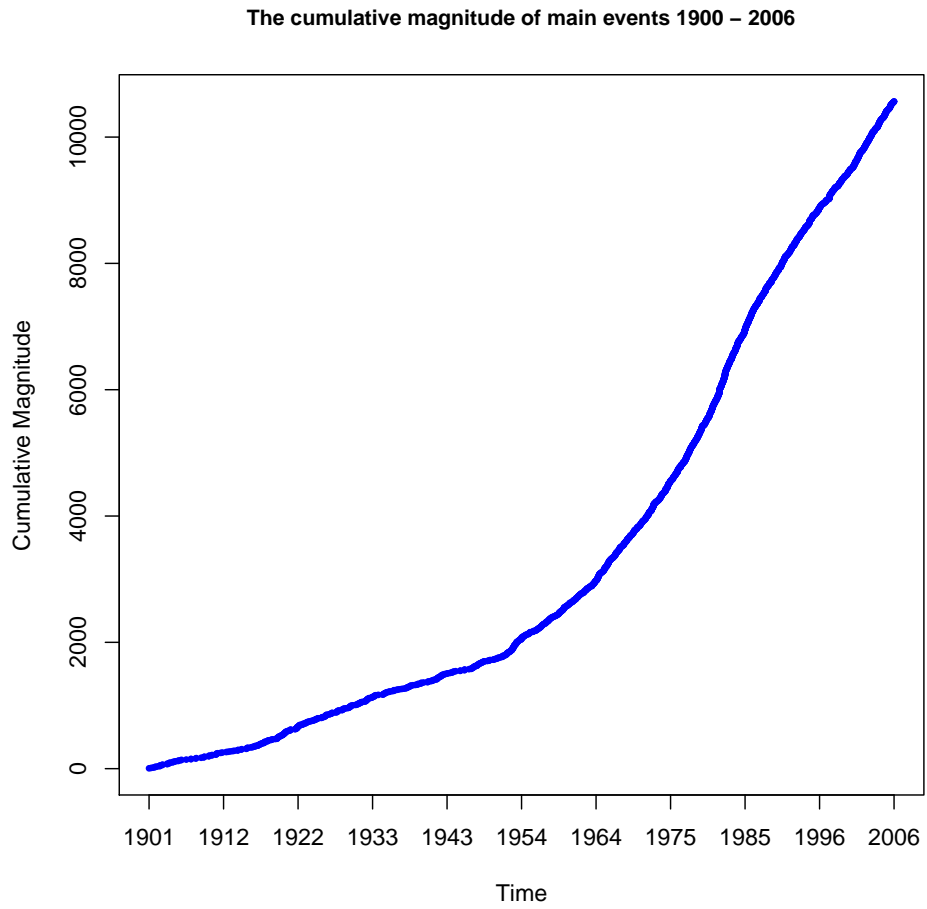}
\vspace{0.5in}
\end{center}
\end{figure}

\begin{figure}
\begin{center}
\caption{Inflection point in Cumulative Magnitude in Greece, 1900 - 2006}\label{fig:04}
\vspace{0.5in}  \includegraphics[width=13cm,height=13cm]{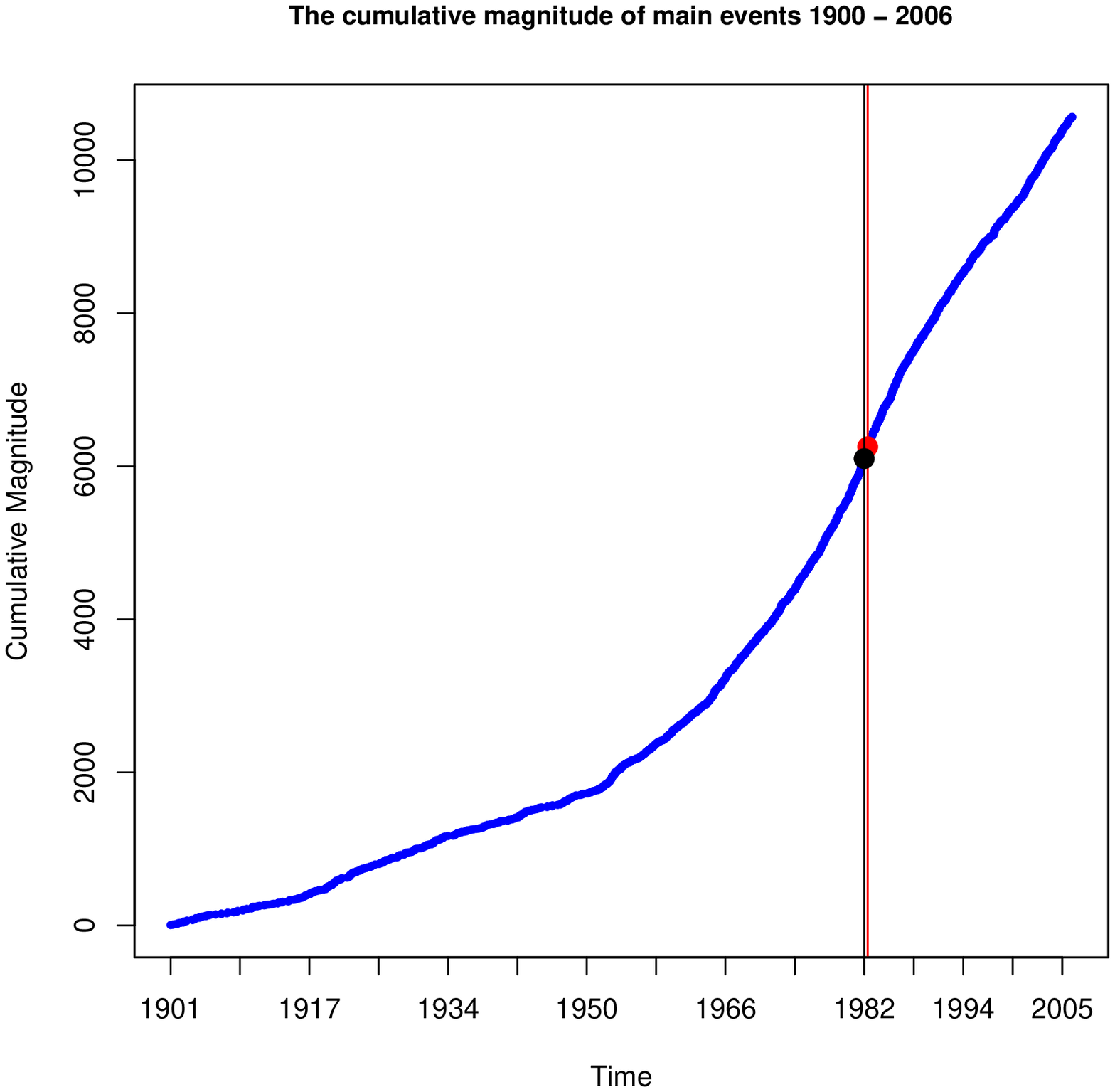}
\vspace{0.5in}
\end{center}
\end{figure}

In order to examine the cumulative energy that is released from main events we just simply compute the cumulative sum of the Magnitudes which is a good index for it. If we plot it versus time we observe that it has an interesting shape, see Figure \ref{fig:03}. \\
We want to investigate which time is the inflection point in this plot in order to see if there exists a significant increase in the released seismic energy. For this purpose we are going to use the methods ESE \& EDE, as described theoretically in ~\cite{dch-12} and as implemented in R Package \textit{inflection}, see ~\cite{dch-13}. The ESE iterations are at Table \ref{tab:02} while the EDE iterations are in Table \ref{tab:03} and the relevant plot at Figure \ref{fig:04}.
\begin{table}[ht]
 \caption{ESE iterations for inflection point of cumulative magnitude} 
    \label{tab:02} 
\centering
\begin{tabular}{rrrrrr}
  \hline
 & \#1 & \#2 & \#3 & \#4 & \#5 \\ 
  \hline
ESE iters & 1992.10 & 1981.22 & 1983.56 & 1982.66 & 1983.11 \\ 
   \hline
\end{tabular}
\end{table}

\begin{table}[ht]
\caption{EDE iterations for inflection point of cumulative magnitude} 
    \label{tab:03}
\centering
\begin{tabular}{rrrrr}
  \hline
 & \#1 & \#2 & \#3 & \#4 \\ 
  \hline
EDE iters & 1985.63 & 1984.48 & 1983.89 & 1982.69 \\ 
   \hline
\end{tabular}
\end{table}

\begin{table}[ht]
\caption{ESE \& EDE first application for cumulative magnitude} 
    \label{tab:04}
\centering
\begin{tabular}{rrrr}
  \hline
 & $i_1$ & $i_2$ & $\chi$ \\ 
  \hline
ESE & 1977.37 & 2006.84 & 1992.10 \\ 
  EDE & 1964.26 & 2007.00 & 1985.63 \\ 
   \hline
\end{tabular}
\end{table}
\clearpage

 If we focus on the intervals that locate the inflection point from the first application of our methods we see from Table \ref{tab:04} that we have a change from mid '60's to mid '70's which is also directly observable at Figures \ref{fig:03} \& \ref{fig:04}. The fact that the second time is exactly the end of time interval on consideration means that we have not a complete sigmoid curve yet.\\
From the mass-energy equivalence in Mechanics we could treat the released energy (as measured by magnitude $M_{i}$) of a main event locally to represent an equivalent of a mass. By taking into account the depth $d_{i}$ of the event we can compute the quantities:
\begin{equation}
\begin{matrix}
I^{(0)}=M_{i}\,d_{i}^{0} \\
I^{(1)}=M_{i}\,d_{i}^{1} \\
I^{(2)}=M_{i}\,d_{i}^{2} \\
\vdots \\
I^{(n)}=M_{i}\,d_{i}^{n} \\
\end{matrix}
\end{equation}
We can call the above quantities `\emph{generalised inertia moments}' (GIM) of the seismic events, since for $i=2$ it is just the inertia moment of a mass that rotates around an axis in a distance $d_{i}$. We continue by computing the relative cumulative GIM for $n=0,1,2,\ldots\,5$. The descriptive statistics of Magnitude and Depth that have been used is presented at Table \ref{tab:05}

\begin{figure}
\begin{center}
\caption{Normalised cumulative $I^{(n)},n=0,1,\ldots,5$ in Greece, 1900 - 2006}\label{fig:05}
\vspace{0.5in}  \includegraphics[width=13cm,height=13cm]{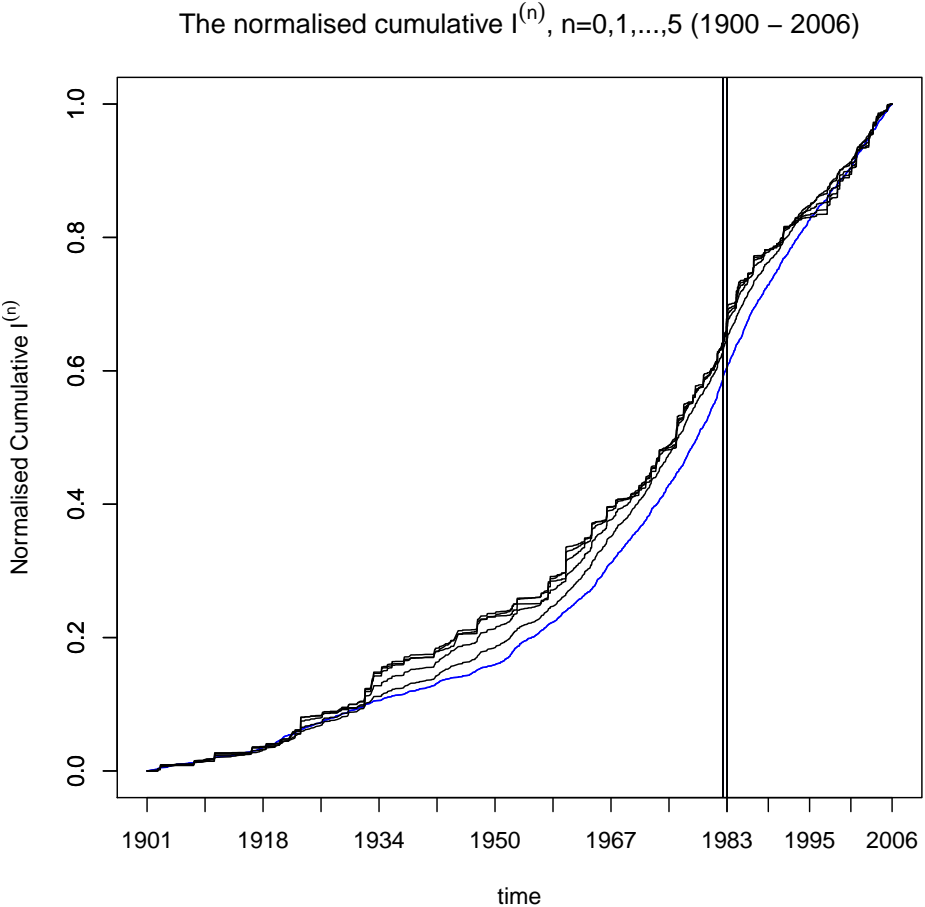}
\vspace{0.5in}
\end{center}
\end{figure}

\begin{figure}[ht]
\begin{center}
\caption{Normalised cumulative $TI^{(n)},n=0,1,\ldots,5$ in Greece, 1900 - 2006}\label{fig:06}
\vspace{0.5in}  \includegraphics[width=10cm,height=10cm]{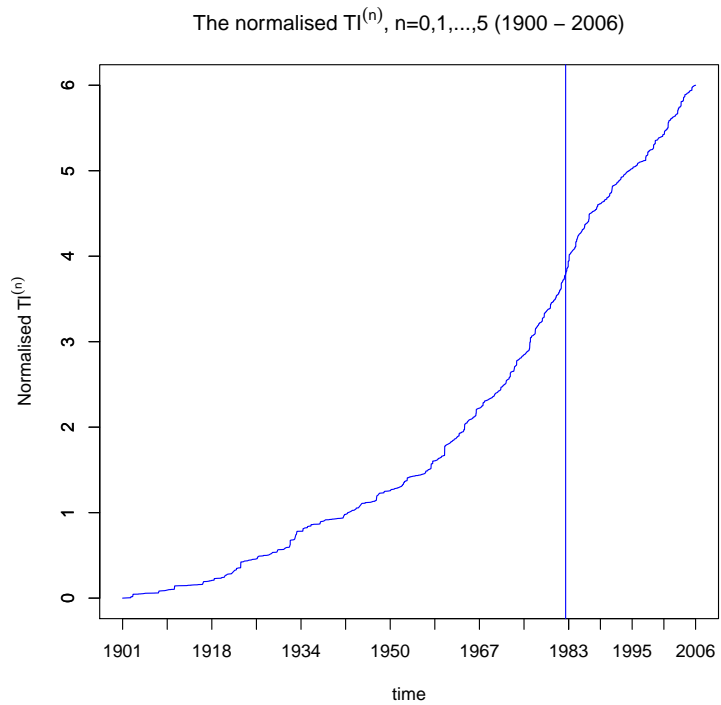}
\vspace{0.5in}
\end{center}
\end{figure}

\begin{table}[!htbp] \centering 
  \caption{Descriptive statistics of Magnitude and Depth, main events, 1900-2006} 
  \label{tab:05} 
\begin{tabular}{@{\extracolsep{5pt}}lccccc} 
\\[-1.8ex]\hline 
\hline \\[-1.8ex] 
Statistic & \multicolumn{1}{c}{N} & \multicolumn{1}{c}{Mean} & \multicolumn{1}{c}{St. Dev.} & \multicolumn{1}{c}{Min} & \multicolumn{1}{c}{Max} \\ 
\hline \\[-1.8ex] 
Magnitude & 2,188 & 4.737 & 0.583 & 4.000 & 7.700 \\ 
Depth & 2,188 & 33.744 & 31.800 & 1 & 199 \\ 
\hline \\[-1.8ex] 
\normalsize 
\end{tabular} 
\end{table} 
We plot now the GIM at Figure \ref{fig:05}, after normalising to the interval $[0,1]$ for comparison reasons. At the same plot we have marked the iterative ESE inflection point. We see that all inflection points are identical and very close to that of the single cumulative magnitude of Figure \ref{fig:04}. \\
\clearpage
Let's try now to add all GIM and obtain the Total GIM (TGIM) for each event
\begin{equation}
TI^{(n)}=\sum_{i=0}^{n}{I^{(n)}}
\end{equation}
We are computing the TGIM and the relevant inflection point of it, after normalising to unity. Results are plotted at Figure \ref{fig:06}. We don't see any remarkable deviation for the inflection point.
\clearpage

\section{Discussion}
Although the magnitude of an earthquake is a measure of the deformation we tried to investigate the situation by introducing the `\emph{generalised inertia moments}' for every event. Our motivation was that of a Taylor series expansion: \textit{Why not to consider all the powers of a reasonable series like that which contains the moment of inertia}? By using iterative ESE \& EDE methods we observed that a time inflection point exist at the end of 1982 to the beginning of 1983. Further investigation is needed for the use of cumulative quantities in describing seismological data.

\bibliographystyle{plain}

\end{document}